\documentclass[notoc]{
JHEP3}
\usepackage{graphicx}

\title{Comment on and Erratum to\\ {}``Pressure of Hot QCD at Large $ N_{f} $''
}

\author{Andreas Ipp \\
Institut f\"ur Theoretische Physik, Technische Universit\"at Wien,\\
  Wiedner Hauptstr. 8--10, A-1040 Vienna, Austria}

\author{Guy D. Moore\\
Department of Physics, McGill University,\\
3600 rue University, Montr\'eal, QC H3A 2T8, Canada}

\author{Anton Rebhan\\
Institut f\"ur Theoretische Physik, Technische Universit\"at Wien,\\
  Wiedner Hauptstr. 8--10, A-1040 Vienna, Austria}

\abstract{%
We repeat and correct
the recent calculation of the thermodynamic potential of hot QCD in
the limit of large number $N_f$ of fermions.
The new result for the thermal pressure
turns out to agree significantly better with results obtained from perturbation
theory at small coupling. For large coupling, a nonmonotonic
behaviour is reproduced, but the pressure of the strongly coupled theory
does not exceed the free pressure
as long as the Landau pole ambiguity remains negligible
numerically.
}

\preprint{TUW-02-26}

\keywords{1/N Expansion, Thermal Field Theory, QCD}

\begin{document}

\section{Introduction}

In a recent paper \cite{Moore:2002md} the thermal
pressure in QCD with a large number of fermions $ N_{f}\gg N_{c}\sim 1 $
was calculated
at next-to-leading order (NLO) in a large $ N_{f} $ expansion.
Although the large-$N_f$ limit is afflicted by the presence of
a Landau pole, thermal effects can be studied in a cutoff theory
provided the temperature is much smaller than the cutoff which
in turn has to be smaller than the scale of the Landau pole.
Then at NLO order of the large $ N_{f} $ expansion 
exact results, nonperturbative in the effective
coupling $g^2_{\rm eff}=g^2N_f/2$, can be obtained 
as long as $g^2_{\rm eff}\ll 6\pi^2$.

Exact large-$N$ results in scalar field theory at finite temperature
have been obtained previously and used to study the
(poor) convergence properties of thermal perturbation theory
\cite{Drummond:1997cw,Bodeker:1998an}.
An exact nonperturbative result for a more QCD-like theory
is of particular interest in view of the various recent attempts to
improve thermal perturbation theory in hot QCD by selective resummations
\cite{Andersen:1999fw,Andersen:2002ey,%
Blaizot:1999ip,%
Peshier:2000hx,Romatschke:2002pb}, for which it may serve
as a testing ground. In Ref.~\cite{Moore:2002md}, it was
proposed to interpret a failure of some technique
at large $N_f$ and reasonably large $g^2_{\rm eff}$ as meaning
that the technique is certainly not valid in full, small-$N_f$ QCD.
However, Peshier \cite{Peshier:2002fm} recently argued that
the strong-coupling behaviour of large $N_f$ QCD is probably
too different from that of small-$N_f$ QCD to draw such
conclusions.

The result presented in Ref.~\cite{Moore:2002md}
is in fact very different from an ideal quasiparticle picture
as pursued in Refs. 
\cite{Peshier:1996ty,Levai:1997yx,Peshier:1999ww}.\footnote{A discussion
of the HTL-quasiparticle picture of QCD thermodynamics underlying
the approach of Ref.~\cite{Blaizot:1999ip} in
the context of large-$N_f$ is contained in Ref.~\cite{Heidelberg}.}
According to Ref.~\cite{Moore:2002md}, the gluonic contribution
decreases as a function of $g^2_{\rm eff}$ only up to
a certain value of $g^2_{\rm eff}$, after which it rises
and even exceeds the free pressure long before the
coupling is so strong that the presence of a Landau pole becomes relevant.

In the following, we shall present the numerical result
that two of us (A.I.~and A.R.) have obtained by a new implementation which
closely follows the approach
of Ref.~\cite{Moore:2002md}. 
This result differs from that published by one
of us (G.D.M.) in Ref.~\cite{Moore:2002md},
but after correcting the error
in the computer code\footnote{In evaluating Eq.~(3.11) of
Ref.~\cite{Moore:2002md}, the imaginary part of the
logarithm of the longitudinal propagator was calculated
as the arctangent of the imaginary part over the real part
without checking whether the argument was within the
principal branch of the arctangent function.} underlying the latter, 
the two independent evaluations agree to an accuracy better than $1:10^{4}$.

The new result turns out to follow rather closely the perturbative
results to order $g^5$ up to $g^2_{\rm eff}\approx 6$.
At $g^2_{\rm eff}\approx 12$ the pressure goes through a minimum
after which it rises, in qualitative accordance
with the result presented in Ref.~\cite{Moore:2002md}, 
but the exact result starts
to exceed the free-gluon pressure only at values of
$g^2_{\rm eff}>28$, which is so large that the Landau pole starts to
influence the results noticeably.

\FIGURE[t]{%
\includegraphics{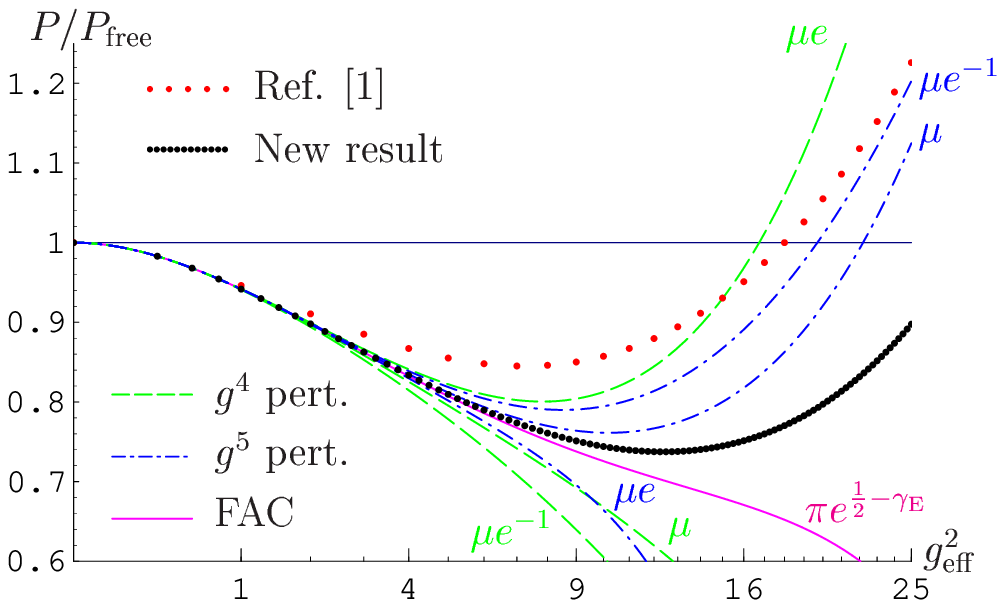}
\centerline{(a)}\\[12pt]
\includegraphics{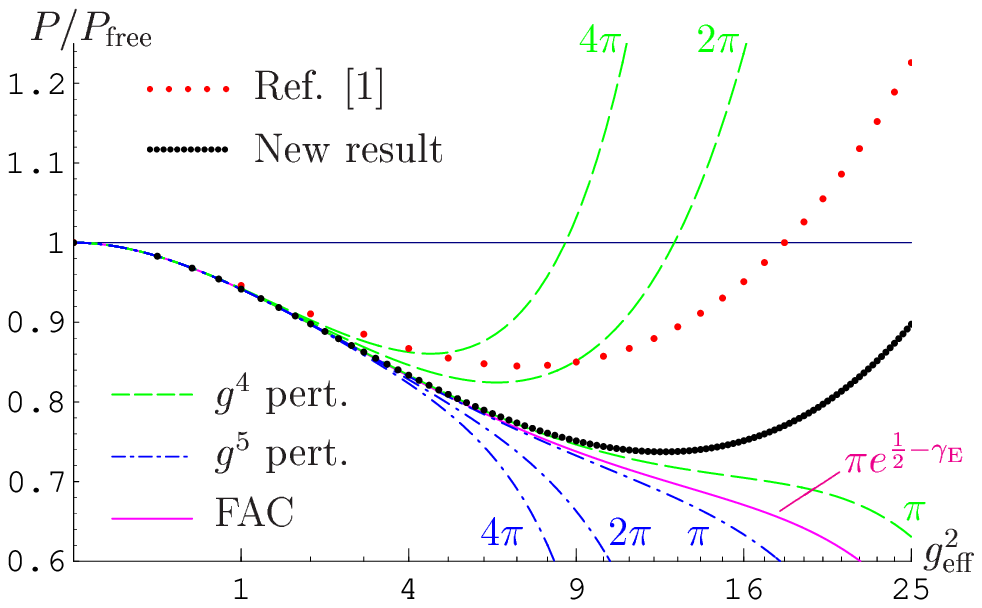}
\centerline{(b)}
\caption{Exact result for \protect$ P_{NLO}/P_{\rm free}\protect $
as a function of $g^2_{\rm eff}(\bar\mu=\pi e^{-\gamma_E}T)$,
rendered with an abscissa linear in $g_{\rm eff}$,
in comparison with the previous result of Ref.~\cite{Moore:2002md}
and two sets of perturbative results through
order \protect$ g^{4}\protect $ and \protect$ g^{5}\protect $:
(a) with renormalization point chosen within a power of $e$
of $\pi e^{-\gamma_E}T$; (b) within a power of 2 of $2\pi T$.
The line marked ``FAC'' corresponds 
$\bar\mu=\pi e^{{1\over2}-\gamma_E}T$
where the perturbative result to order $g^4$ coincides with
the one to order $g^5$.
\label{figurePnlo}}
}

\section{Results}

The NLO contribution to the thermal pressure, of order $N_f^0$, is given by
the one-loop gauge-boson contribution with
any number of (renormalized) fermion bubble insertions \cite{Moore:2002md}.
Carrying out the sums over Matsubara frequencies, this
expression involves terms proportional to the Bose distribution $n_b$,
which are best calculated in Minkowski space, and parts
without this factor, which Ref.~\cite{Moore:2002md}
evaluated partly in Minkowski and partly in Euclidean space.
To avoid spurious logarithmic divergences, it is crucial
to employ a Euclidean invariant cutoff $\Lambda$ when cutting out
the Landau pole. This introduces an error which is suppressed,
relative to the full thermal contribution, by
$\sim T^4/\Lambda^4$, so that the ambiguity
caused by the Landau singularity is well under control for $\Lambda\gg T$.
If the coupling $g^2_{\rm eff}\ll 6\pi^2$, the Landau
pole is exponentially large and one may choose a large
cutoff $\Lambda^2=a \Lambda^2_{\rm Landau}$, which
following \cite{Moore:2002md} we shall vary by taking $a$ between $1/4$
and $1/2$.

To ensure Euclidean O(4) invariance when performing
parts of the calculation in Minkowski and parts in
Euclidean space, which have to be connected by great arcs,
one needs the analytic continuation
of the complete fermion one-loop self-energy to the
complex energy plane. 
The relevant formulae are listed in the Appendix.

Ref.~\cite{Moore:2002md} calculated pieces linear in $ n_{b} $
in Minkowski space.  Terms without $ n_{b} $ were computed along a
complex frequency contour which ran up the Minkowski axis 
to $ \omega_{\rm max}<\Lambda_{\rm Landau}\sqrt{a} $ for some $ a<1 $,
then along the great arc to Euclidean space, and back down
to $ q_{0}=\sqrt{q^{2}_{max}-q^{2}} $; finally, a Euclidean integration
of the $n_{b}$ free term was performed
over 4-spheres in Euclidean space up to $ Q^{2}<\Lambda ^{2}_{\rm Landau}a $. 

It is in fact simpler to calculate all pieces linear in $ n_{b} $
in Minkowski space, and all terms without $ n_{b} $ in Euclidean space.
By actually calculating both ways, we
have a rather non-trivial numerical check on the result. 
In our numerical implementation both ways turned out to 
agree within numerical errors of about
$ 10^{-5} $.

In Fig.~\ref{figurePnlo} we give our numerical result as a function
of $g^2_{\rm eff}(\bar\mu=\pi e^{-\gamma_E}T)$.\footnote{Tabulated
results can be obtained on-line from
\href{http://hep.itp.tuwien.ac.at/~ipp/data/}{\tt 
http://hep.itp.tuwien.ac.at/~{}ipp/data/} .}
The new result agrees well with the perturbative results to order $g^5$
up to $g^2_{\rm eff}\approx 5$, where the renormalization scheme
dependence\footnote{In contrast to the exact result, the perturbative
results depend on the value of the renormalization point $\bar\mu$,
which we vary between $\pi T$ and $4\pi T$, expressing everything
as a function of $g^2_{\rm eff}(\bar\mu=\pi e^{-\gamma_E}T)$, however,
to make a comparison possible.}
of the $g^5$-result is still reasonably small (the previous
result of Ref.~\cite{Moore:2002md} showed significant deviations from
the perturbative results already for $g^2_{\rm eff}\gtrapprox 2$).
If the perturbative result to order $g^5$ is optimized by
fastest apparent convergence (FAC), which requires that
the result to order $g^4$ coincides with the one to order
$g^5$ and which amounts to
$\bar\mu=\pi e^{{1\over2}-\gamma_E}T$, the agreement
with perturbation theory is improved and extends to 
$g^2_{\rm eff}\approx 7$.

For higher values of $g^2_{\rm eff}$ the exact result flattens out
and reaches a minimum at $g^2_{\rm eff}\approx 12$. For still higher
values the pressure rises but,
contrary to the previous result of \cite{Moore:2002md}, it
does not exceed the free pressure
for the range of coupling considered in \cite{Moore:2002md}.

In Fig.~\ref{figurePland} we consider even higher values
of $ g_{\rm eff}^{2} $ and find that eventually
the thermal pressure grows larger than the free pressure.
This occurs at $g_{\rm eff}^{2} > 28$ where
$\Lambda_{\rm Landau}/T < 34$. While this still seems to be
a reasonably large number, the numerical result starts to
become sensitive to the cutoff just where the pressure
approaches the free one. 
The four curves displayed in
Fig.~\ref{figurePland} show the result of varying the parameter $a$
in the UV cutoff $\sqrt{a}\Lambda_{\rm Landau}$ in the
Minkowski and Euclidean parts of the calculation ($a_M$ and
$a_E$, resp.) from $a=1/4$ to $a=1/2$.
The numerical result is rather insensitive to this below
$g^2_{\rm eff}\approx 25$, but very sensitive in
the region where the pressure starts to exceed the free one.

\FIGURE[t]{%
\includegraphics{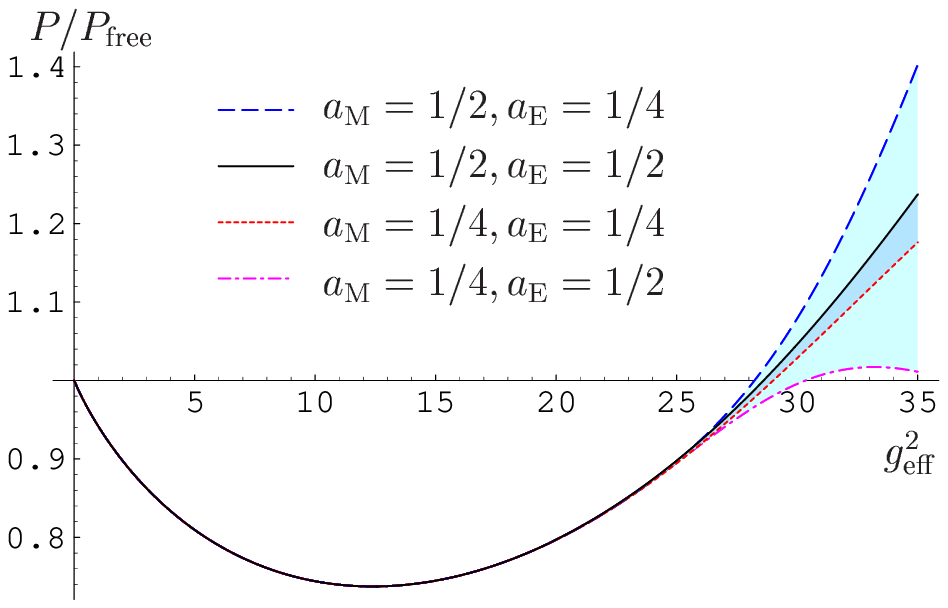}
\caption{The result for \protect$ P_{NLO}/P_{\rm free}\protect $
up to $g_{\rm eff}^{2}=35$ and 
with different cutoffs.
\label{figurePland}}
}

\section{Conclusion}

The exact result for the pressure of hot QCD in the limit of
large $N_f$ 
shows a nonmonotonic behaviour as a function of
the coupling. The minimum of the pressure is reached
when $\Lambda_{\rm Landau} \simeq 480 T$ and where the
ambiguity introduced by the Landau singularity is
completely negligible. For higher values of the coupling
the pressure eventually reaches and exceeds that of the
free theory, but at that point the Landau pole
is at $\Lambda_{\rm Landau} < 34 T$. Above this
point the result becomes increasingly sensitive to the precise
cutoff which has to be chosen 
between $T$ and $\Lambda_{\rm Landau}$. This suggests
that only the nonmonotonic behaviour 
is to be taken seriously, but not the fact that
the free theory value is eventually reached and exceeded.

So in contrast to the previous result
of \cite{Moore:2002md}, the corrected one
does not imply that an ideal quasiparticle picture
(where the pressure has to be smaller than the free one)
is necessarily in conflict with the actual physics of QCD in the limit
of large fermion number. In order to be compatible with
the nonmonotonic
behaviour at large coupling, 
however, a quasiparticle picture 
would require a correspondingly nonmonotonic behaviour
of the quasiparticle masses. While this is not
particularly natural for the simple quasiparticle picture
underlying the approaches of 
\cite{{Peshier:1996ty,Levai:1997yx,Peshier:1999ww}}, this is not
a priori
excluded for the more complicated HTL-based ones of
Ref.~\cite{Blaizot:1999ip}.
This issue is discussed in more detail in Ref.~%
\cite{Heidelberg}.

\appendix
\section{Appendix: Gauge-boson self-energy}

\subsection{Spectral representation}

A convenient 
starting point for performing the analytic continuation of the self-energy
from Minkowski to Euclidean space or vice versa is its spectral representation
\begin{equation}
\label{spectralform1}
\Pi _{\mu \nu }(z,\mathbf{p})=
\int _{-\infty }^{\infty }
\frac{dp_{0}}{2\pi }\frac{\pi _{\mu \nu }(p_{0},p)}{z-p_{0}}
\end{equation}
which is valid for any complex $ z $ \cite{Landsman:1987uw}.
The spectral form $\pi _{\mu \nu }(p_{0},p)$ is a purely real
quantity that can be read off from the fermion loop evaluated
in the imaginary time formalism according to
\begin{eqnarray}
\Pi ^{\mu \nu }(i\omega _{n},\mathbf{p}) & = & -4g^{2}_{\rm eff}\int \frac{d^{3}k}{(2\pi )^{3}}\int _{-\infty }^{\infty }\frac{dk_{0}}{2\pi }\int _{-\infty }^{\infty }\frac{dq_{0}}{2\pi }\label{selfenergyimaginarytime} \\
 &  & \times \rho _{0}(k_{0},\mathbf{k})\rho _{0}(q_{0},\mathbf{q})
\frac{n_f(k_{0})-n_f(q_{0})}{k_{0}-q_{0}-i\omega _{n}}I^{\mu \nu }(k,q)\nonumber 
\end{eqnarray}
with $ I^{\mu \nu }=k^{\mu }q^{\nu }+q^{\mu }k^{\nu }
-g^{\mu \nu }k^{\alpha }q_{\alpha }+g^{\mu \nu }m^{2} $
and $ \mathbf{q}\equiv \mathbf{k}-\mathbf{p} $. The fermionic distribution 
function
is given by $ n_f(k_{0})=1/(e^{k_{0}/T}+1) $ and the free spectral function
by $ \rho _{0}(k_{0},\mathbf{k})=2\pi \epsilon (k_{0})\delta (k_{0}^{2}-\mathbf{k}^{2}-m^{2})=\frac{\pi }{\varepsilon _{k}}(\delta (k_{0}-\varepsilon _{k})-\delta (k_{0}+\varepsilon _{k})) $
with $ \epsilon (k_{0})=k_{0}/|k_{0}| $ and $ \varepsilon _{k}\equiv \sqrt{\mathbf{k}^{2}+m^{2}} $.
(In the following we shall however consider only the ultrarelativistic
limit $m/T\to 0$.)

We need separately the transverse and longitudinal projection of
the self-energy. Following Weldon \cite{Weldon:1982aq} we define
$ 2g_{\rm eff}^{2}G\equiv g^{\mu \nu }\Pi _{\mu \nu } $ and $ 2g^{2}_{\rm eff}H\equiv u^{\mu }u^{\nu }\Pi _{\mu \nu } $
with the thermal rest frame velocity $ u^{\mu }=(1,0,0,0) $.

Treating
the various projections of the spectral density separately, 
we obtain the following useful representations
by analytically performing three of the four integrations: 
\begin{equation}
\label{spectraldensityantisym1}
\pi _{X}(p_{0},\mathbf{p})=\pi _{X}^{+}(p_{0},p)-\pi _{X}^{+}(-p_{0},p)
\end{equation}
 \begin{eqnarray}
\pi _{X}^{+}(p_{0},p) & = & \frac{g^{2}N_{f}}{2\pi p}\int _{0}^{\infty }dk
\left( n_f(k)-\frac{1}{2}\right) \bar{I}_{X}\\
 &  & \times \epsilon (k-p_{0})\theta (|k-p|\leq |k-p_{0}|\leq |k+p|)\nonumber 
\end{eqnarray}
where the $ \theta  $-function stems from the angular integration
between $ \mathbf{p} $ and $ \mathbf{k} $ (its usage here means
$ \theta (\textrm{true expression})=1 $ and $ \theta (\textrm{false expression})=0 $)
and $ X $ = $ G $ or $ H $ as in\begin{eqnarray}
\bar{I}_{G}\, \equiv \, g^{\mu \nu }I_{\mu \nu }(k_{0}=k) & = & p_{0}^{2}-p^{2}\\
\bar{I}_{H}\, \equiv \, u^{\mu }u^{\nu }I_{\mu \nu }(k_{0}=k) & = & \frac{1}{2}(2k+p-p_{0})(2k-p-p_{0}).
\end{eqnarray}
The spectral density $ \pi  $
is manifestly real and odd in $ k_{0} $, i.e. 
$ \pi (k_{0},\mathbf{k})=-\pi (-k_{0},\mathbf{k}). $ 

To subtract the vacuum part, one just has to 
replace $ \left( n_f(k)-\frac{1}{2}\right)  $
by $ n_f(k) $. We shall do so in the following explicit results,
because the vacuum part requires regularization and renormalization,
after which the (Euclidean) self energy simply reads
\begin{equation}
\Pi_{\rm vac.}^{\mu\nu}=
-{g^2_{\rm eff}\over 12\pi^2}\left(\eta^{\mu\nu}P^2-{P^\mu P^\nu}\right)
\left(\ln{P^2\over \bar\mu^2}-{5\over 3}\right).
\end{equation}

\subsection{Minkowski result} 

For Minkowski space we use the Feynman prescription\footnote{%
Note that with our expressions one has to turn the
Euclidean $ p_{0} $ into the lower half of the complex plane $ p_{0}\rightarrow -i\omega +\epsilon  $
to obtain the retarded self-energy.}
$ \widetilde{\Pi }^{\rm F}(p_{0},\mathbf{p})\equiv \Pi (p_{0}+ip_{0}\varepsilon ,\mathbf{p}) $
for which the self-energy can be separated into \begin{equation}
\label{selfenergyfeynmanreal}
\textrm{Re}\widetilde{\Pi }^{\rm F}(p_{0},\mathbf{p})=\int _{-\infty }^{\infty }\frac{dp'_{0}}{2\pi }\pi (p'_{0},p)\frac{\textrm{P}}{p_{0}-p'_{0}}
\end{equation}
\begin{equation}\label{selfenergyfeynmanimag}
\textrm{Im}\widetilde{\Pi }^{\rm F}(p_{0},\mathbf{p})=-\frac{1}{2}\epsilon (p_{0})\pi (p_{0},p).
\end{equation}
 with $ \textrm{P} $ denoting the principal value as in $ \frac{1}{x+i\varepsilon }=\frac{\textrm{P}}{x}-i\pi \delta (x) $.

Inserting (\ref{spectraldensityantisym1}) in the expressions
(\ref{selfenergyfeynmanreal}) 
and (\ref{selfenergyfeynmanimag})
we reproduce the real part of the
self-energy as given in Weldon's paper \cite{Weldon:1982aq}
\begin{equation}
\textrm{Re}\widetilde{\Pi }_{G}(p_{0},\mathbf{p})  =  \frac{g^{2}N_{f}}{2\pi ^{2}}\int _{0}^{\infty }dk \;n_f(k) \left[ 4k+\frac{p_{0}^{2}-p^{2}}{2p}\log \left| \frac{2k+p_{0}+p}{2k+p_{0}-p}\frac{2k-p_{0}+p}{2k-p_{0}-p}\right| \right] \nonumber 
\end{equation}
and\begin{eqnarray}
\textrm{Re}\widetilde{\Pi }_{H}(p_{0},\mathbf{p}) & = & \frac{g^{2}N_{f}}{2\pi ^{2}}\int _{0}^{\infty }dk \;n_f(k) \left[ 2k\left( 1-\frac{p_{0}}{p}\log \left| \frac{p_{0}+p}{p_{0}-p}\right| \right) \right. \nonumber \\
 &  & \quad \; \left. +\frac{(2k+p_{0}+p)(2k+p_{0}-p)}{4p}\log \left| \frac{2k+p_{0}+p}{2k+p_{0}-p}\right| \right. \nonumber \\
 &  & \quad \; \left. -\frac{(2k-p_{0}-p)(2k-p_{0}+p)}{4p}\log \left| \frac{2k-p_{0}-p}{2k-p_{0}+p}\right| \right] .
\end{eqnarray}


The imaginary part was not explicitly calculated by Weldon, but 
we can provide a completely
analytical result where no integration is left to be performed. It
is given by 

\begin{eqnarray}
\textrm{Im}\widetilde{\Pi }_{X}(p_{0},\mathbf{p}) & = & -\frac{1}{2}\epsilon (p_{0})\frac{g^{2}N_{f}}{2\pi p}\left[ F^{S}_{X}(\frac{\left| p_{0}+p\right| }{2})-F^{S}_{X}(\frac{\left| p_{0}-p\right| }{2})\right. \nonumber \\
 &  & \qquad \left. +\epsilon (p_{0}+p)F^{A}_{X}(\frac{\left| p_{0}+p\right| }{2})-\epsilon (p_{0}-p)F^{A}_{X}(\frac{\left| p_{0}-p\right| }{2})\right] 
\end{eqnarray}
with symmetric and antisymmetric functions $ F_{X}^{S}\equiv (F_{X}^{+}+F_{X}^{-})/2 $
and $ F_{X}^{A}\equiv (F_{X}^{+}-F_{X}^{-})/2 $ that are defined
as \begin{equation}
F^{\pm }_{G}(x)\equiv \int _{x}^{\infty }\;n_f(k)\bar{I}_{G}(\pm p_{0},p,k)dk=(p_{0}^{2}-p^{2})F_{1}(x)
\end{equation}
\[
F^{\pm }_{H}(x)\equiv \int _{x}^{\infty }\;n_f(k)\bar{I}_{H}(\pm p_{0},p,k)dk=\frac{p_{0}^{2}-p^{2}}{2}F_{1}(x)\mp 2p_{0}F_{2}(x)+2F_{3}(x),\]
where the $ F_{i}(x) $ are the following integrals 
\begin{eqnarray}
F_{1}(x) & \equiv  & \int _{x}^{\infty }\,n_f(k)dk=-x+T\log (e^{x/T}+1),\\
F_{2}(x) & \equiv  & \int _{x}^{\infty }k\,n_f(k)dk=\frac{\pi ^{2}T^{2}}{6}-\frac{x^{2}}{2}+xT\log (e^{x/T}+1)+T^{2}\textrm{Li}_{2}(-e^{x/T}),\\
F_{3}(x) & \equiv  & \int _{x}^{\infty }k^{2}\,n_f(k)dk=-\frac{x^{3}}{3}+x^{2}T\log (e^{x/T}+1)\nonumber\\
 &  & \qquad \qquad \qquad \quad \, \, +2xT^{2}\textrm{Li}_{2}(-e^{x/T})-2T^{3}\textrm{Li}_{3}(-e^{x/T}),
\end{eqnarray}
with $ \textrm{Li}_{n}(x) $ being the polylogarithm function. Note
that $ F_{G}^{A}=0 $ simplifies our expression for $ \textrm{Im}\widetilde{\Pi }_{G} $
considerably. 

\subsection{Euclidean result}

For Euclidean space we set $ z=i\omega  $ and (using the antisymmetry
property of the spectral density) we get
\begin{equation}
\textrm{Re}\widetilde{\Pi }^{\rm Eucl}(i\omega ,\mathbf{p})=\int _{-\infty }^{\infty }\frac{dp_{0}}{2\pi }\pi (p_{0},p)\frac{-p_{0}}{\omega ^{2}+p_{0}^{2}},
\qquad
\textrm{Im}\widetilde{\Pi }^{\rm Eucl}(i\omega ,\mathbf{p})=0.
\end{equation}

We are left with real integrals of the 
form
\begin{equation}
\int dp_{0}\frac{-2p_{0}}{\omega ^{2}+p_{0}^{2}}\bar{I}_{G}(p_{0},p,k)=
-p_{0}^{2}+(\omega ^{2}+p_{0}^{2})\log (\omega ^{2}+p_{0}^{2})
\end{equation}
and\begin{eqnarray}
\int dp_{0}\frac{-2p_{0}}{\omega ^{2}+p_{0}^{2}}\bar{I}_{H}(p_{0},p,k) & = & \frac{1}{2}p_{0}(8k-p_{0})-4k\omega \arctan \left( \frac{p_{0}}{\omega }\right) \nonumber \\
 &  & -\frac{1}{2}(4k^{2}-p^{2}-\omega ^{2})\log (\omega ^{2}+p_{0}^{2}).
\end{eqnarray}
With the appropriate integration limits
we finally obtain
the self-energy in Euclidean space as
\begin{eqnarray}\label{euclideang}
\textrm{Re}\widetilde{\Pi }_{G}(i\omega ,\mathbf{p}) & = & \frac{g^{2}N_{f}}{2\pi ^{2}}\int _{0}^{\infty }dk \;n_f(k) \left( 4k+\frac{\omega ^{2}+p^{2}}{2p}\log \frac{\omega ^{2}+(2k-p)^{2}}{\omega ^{2}+(2k+p)^{2}}\right) \\ 
\label{euclideanh}
\textrm{Re}\widetilde{\Pi }_{H}(i\omega ,\mathbf{p}) & = & \frac{g^{2}N_{f}}{2\pi ^{2}}\int _{0}^{\infty }dk \;n_f(k) \label{selfenergyeuclidresultH} 
\left[ 2k+\frac{\omega ^{2}+p^{2}-4k^{2}}{4p}\log \frac{\omega ^{2}+(2k-p)^{2}}{\omega ^{2}+(2k+p)^{2}}\right. \nonumber \\
 &  & \, \, \left. -\frac{2k\omega }{p}\left( \arctan \frac{2k-p}{\omega }+2\arctan \frac{p}{\omega }-\arctan \frac{2k+p}{\omega }\right) \right] .
\end{eqnarray}
This is in principle the result given in the Appendix of \cite{Moore:2002md}
where the three terms involving the arc tangents are replaced by
a common logarithm according to
\begin{equation}\label{arctansintolog}
\arctan \frac{2k-p}{\omega }+2\arctan \frac{p}{\omega }
-\arctan \frac{2k+p}{\omega } = 
-\frac{i}{2}\log \frac{1+\frac{4k^{2}}
{(\omega -ip)^{2}}}{1+\frac{4k^{2}}{(\omega +ip)^{2}}}.
\end{equation}
However, while taking the principal branch of the arctan functions gives
a smooth function over all $ k $, on the right-hand side
one must not restrict to the principal branch of the logarithm.

For verifying the path independence of the numerical results
we also need the self energy for complex energies.
These may be obtained either from the analytic continuation
of the results (\ref{euclideang}) and (\ref{euclideanh}) or
from the spectral representation according to
\begin{equation}
\label{selfenergycomplex2}
\Pi (a+ib,\mathbf{p})=\int \frac{dp_{0}}{2\pi }\pi (p_{0},p)\left( \frac{a-p_{0}}{(a-p_{0})^{2}+b^{2}}-i\frac{b}{(a-p_{0})^{2}+b^{2}}\right),
\end{equation}
with real and unambiguous integrals (for $ b\neq 0 $).

\acknowledgments

This work has been supported by the Austrian Science Foundation FWF,
project no. 14632-TPH.


\providecommand{\href}[2]{#2}
\begingroup\raggedright\endgroup
\end{document}